\documentclass[12pt]{article}
\usepackage[margin=2.5cm]{geometry}
\usepackage{amsmath, amssymb}
\usepackage{color}
\usepackage{float}
\usepackage{graphicx}
\usepackage{subcaption,adjustbox}
\renewcommand{\Re}{\mathrm{Re}}
\renewcommand{\Im}{\mathrm{Im}}

\begin{document}

\title{Geometric Minimization of Softly-Broken Potentials}

\author{
Ivo de Medeiros Varzielas \footnote{ivo.de@udo.edu},
Diogo Ivo \footnote{diogo.ivo@tecnico.ulisboa.pt}\\
CFTP, Departamento de Física,\\
Instituto Superior T\'{e}cnico, Universidade de Lisboa, \\ Avenida Rovisco Pais 1, 1049 Lisboa, Portugal \\[5pt]
}

\maketitle

\begin{abstract}
We study the minimization of multi-Higgs models with symmetries that are softly-broken. The powerful method of geometric minimization enables analytic minimization of multi-Higgs models with large symmetries. When these symmetries are softly-broken, the method needs to be adapted. We propose a useful generalization that considers the effect of the soft-breaking terms to the quadratic part of the potential, by applying the procedure to restricted orbit spaces. We exemplify our novel methodology by finding and classifying the minima for an $S_4$ multi-Higgs model that is softly-broken with specific terms.
\end{abstract}

\section{Introduction}

The Standard Model (SM) is unable to account for the Dark Matter abundance observed in the Universe. This inability is a strong motivation to study theories Beyond the SM. Among these, a very simple class of model considers extending the field content with additional Higgs doublets - Multi-Higgs Doublet Models (MHDMs), which can provide dark matter candidates as well as the possibility of spontaneous CP violation.

Within MHDMs, the simplest generalization is 
the 2HDM \cite{Lee:1973iz}, introduced to provide CP violation. The 2HDM is very well studied (see e.g. the reviews \cite{Branco:2011iw, Ivanov:2017dad}). The 3HDM \cite{Weinberg:1976hu}, introduced also with CP violation in mind, has several features worth considering. Due to the large number of parameters in the most general 3HDM, symmetries are often considered \cite{Ivanov:2012ry, Darvishi:2021txa, deMedeirosVarzielas:2021zqs, deMedeirosVarzielas:2022kbj}.

The minimization of the potential to find the alignment of the vacuum expectation values (vevs) is important but non-trivial and usually becomes impossible to perform analytically. A very powerful method, of Geometric Minimization \cite{Degee:2012sk}, has been proposed to address this problem in the presence of MHDM potentials with large symmetries \footnote{Another method has been proposed in \cite{deMedeirosVarzielas:2017glw}.}. With the great phenomenological interest in softly-broken potentials \cite{deMedeirosVarzielas:2021zqs,deMedeirosVarzielas:2022kbj}, we extend here the method of geometric minimization to the study of the vevs of softly-broken potentials. We present the general considerations and exemplify the method in the potentials of softly-broken $S_4$ 3HDM.

In Section \ref{sec:geo} we review the geometric minimization method and generalize the formalism to apply to potentials with soft-symmetry breaking. A detailed example of the application of this method to the $S_4$ invariant potential follows in Section \ref{sec:example}. The minima are presented in Section \ref{sec:minima} and we present our conclusions in Section \ref{sec:con}.

\section{Geometric Minimization \label{sec:geo}}

\subsection{Fully-Symmetric Models}
We begin by reviewing the geometric method described in \cite{Degee:2012sk} in the context of 3HDM, in preparation of what is to follow.

First, one needs to introduce the set of appropriate variables for the problem. In any given 3HDM the scalar potential is written as a function of 3 Higgs doublets, $\Phi_i$, which for the purposes of this work can be taken as doublets of complex numbers,

\begin{equation}
    \label{eq::doublet}
    \Phi_i =  \begin{pmatrix}
    c_i e^{i\theta_i} \\
    v_i e^{i\omega_i}\\ 
    \end{pmatrix}, \quad i=1, 2, 3.
\end{equation}
We consider then as our variables the following 9 combinations constructed from bilinear contractions of the $\Phi_i$,

\begin{align}
\label{eq::bilinears}
    r_0 & = \frac{\Phi_1^\dagger \Phi_1 + \Phi_2^\dagger \Phi_2 + \Phi_3^\dagger \Phi_3}{\sqrt{3}}, & r_3 & = \frac{\Phi_1^\dagger \Phi_1 - \Phi_2^\dagger \Phi_2}{2}, & r_8 & = \frac{\Phi_1^\dagger \Phi_1 + \Phi_2^\dagger \Phi_2 - 2 \Phi_3^\dagger \Phi_3}{2\sqrt{3}}, \notag \\
    r_1 & = \Re (\Phi_1^\dagger \Phi_2),  & r_4 & = \Re (\Phi_3^\dagger \Phi_1), & r_6 & = \Re (\Phi_2^\dagger \Phi_3), \notag \\
    r_2 & = \Im (\Phi_1^\dagger \Phi_2),  & r_5 & = \Im (\Phi_3^\dagger \Phi_1), & r_7 & = \Im (\Phi_2^\dagger \Phi_3),
\end{align}

\noindent which satisfy

\begin{align}
\label{eq::bilinear_properties}
    r_0 & \geq 0, & \frac{1}{4} r_0 ^2 & \leq \sum_{i = 1}^8 r_i^2 \leq r_0 ^ 2.
\end{align}

\noindent One important aspect of the last expression is that equality is reached, that is, 

\begin{equation}
\label{eq::neutral_equality}
    \sum_{i = 1}^8 r_i^2 = r_0 ^ 2,
\end{equation}

\noindent if and only if one works in a neutral vacuum, with all $c_i = 0$ up to an $SU(2)$ rotation. We also define the variables

\begin{equation}
    y_i = \frac{r_i}{r_0}.
\end{equation} An important property of these is that they are sensitive only to the overall vev alignment. More precisely, given a set of vevs $(v_1,v_2,v_3)$, the same value of $y_i$ would be obtained for any rescaled alignment of the form $\beta (v_1,v_2,v_3), \, \beta > 0$, since both $r_0$ and the $r_i$ are homogeneous functions of degree two in $\beta$.

In this formalism, the most general 3HDM potential is written as

\begin{equation}
\label{eq::genpotential}
    V = - M_0 r_0 - M_i r_i + \Lambda_{00} r_0 ^2 + \Lambda_{0i}r_0 r_i + \Lambda_{ij} r_i r_j,
\end{equation} which has 64 free parameters. However, in the context of 3HDMs (and MHDMs in general), in order to control this proliferation of the number of degrees of freedom it is usual to make $V$ symmetric under some symmetry group $G$. From now on let us consider only symmetries that impose that all $M_i = 0$, so that only $M_0$ remains. This is the case, for example, for $G = \{A_4, S_4, \Delta(54), \Sigma(36)\}$ \cite{Ivanov:2014doa}. As intended, these groups also drastically reduce the number of quartic parameters of the potential, by enforcing certain equalities between themselves, or by setting some of them to be zero.
Let us denote the final number of free quartic parameters by $k+1$, where for the aforementioned groups $k$ does not exceed four. Thus, in these cases we can recast \eqref{eq::genpotential} in a more convenient form,

\begin{equation}
\label{eq::recast_potential}
    V = - M_0 r_0 + r_0 ^2 (\Lambda_0 + \sum_{i=1}^k \Lambda_{i} x_i )= - v_2 r_0 + r_0 ^2 v_4,
\end{equation}

\noindent with

\begin{align}
\label{eq::canonical}
    v_2 & = M_0, & v_4 & = \Lambda_0 + \sum_{i=1}^k \Lambda_{i} x_i.
\end{align}

\noindent Here, the parameter $M_0$ in assumed to be positive. Within $v_4$ the $x_i$ are the combinations of the $y^2$ that, due to the symmetry group, end up coupling to the same parameter $\Lambda_i$ \footnote{The $\Lambda_{i}$ are now functions of the various $\Lambda_{ij}$ in \eqref{eq::genpotential}, whose specific form is dictated by the specific symmetry one works with. We have that $x_0 = 1$ by convention meaning that $\Lambda_{0}=\Lambda_{00}$.}. Note that since the $x_i$ are invariant under rescaling by $\beta$, then so is $v_4$, and as such this recast has led to a separation of the dependence of $V$ on the total magnitude $v^2 = v_1^2 + v_2^2 + v_3^2$ and on the specific alignments compatible with this magnitude, akin to a separation of radial and angular variables in spherical coordinates.

Explicitly, this means that for any continuous set of vevs $\beta (v_1,v_2,v_3), \; \; \beta > 0$ with $v_i$ fixed we will always obtain the same value of $v_4$, and only the value of $r_0$ will change. Thus, from this point of view, for the set $(v_1,v_2,v_3)$, $V$ is a function of $r_0$ only, with minimum

\begin{equation}
    \label{eq::pot_min}
    V_{min}^{(v_1,v_2,v_3)} = - \frac{v_2^2}{4 v_4},
\end{equation}

\noindent occurring at the magnitude

\begin{equation}
    r_0 = \frac{v_2}{2 v_4}.
\end{equation}

Then, the minimization of the potential reduces to determining which alignment leads to the lowest value of $V$. Looking at \eqref{eq::pot_min}, we conclude that the desired alignment is the one that leads to the smallest positive value of $v_4$, since $v_2$ is simply a constant. The positivity of $v_4$ ensures that the model is bounded from below. 

\begin{figure}[H]
\centering
        \input{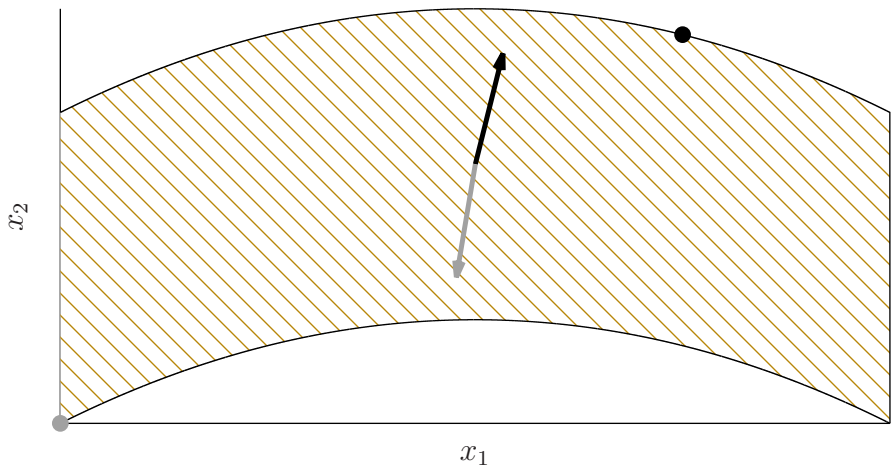}
        \caption{Hypothetical two-dimensional $\Gamma$. In this Figure we display two different vectors $\Vec{n}$ and the point which maximizes $\Vec{n} \cdot \Vec{x}$ in matching colour. Note that for the black vector small changes in the values of $\Lambda_i$ lead to corresponding variations of the coordinates of the point, due to the locally convex geometry of $\Gamma$, whereas for the gray vector the opposite occurs, and the locally concave geometry protects its point against perturbations. However, were this vector to point directly downwards then we would have a doubly degenerate minimum, with both points lying on the $x_1$ axis leading to the same value of the potential.}
        \label{fig::gammaspace}
\end{figure}

In order to determine the minimum alignment we take advantage of the linear nature of $v_4$ on the $x_i$, which reduces the minimization to the determination of the point that minimizes $\Lambda_i x_i$ for a given set of $\Lambda_i$, or, conversely, maximizes the dot product $\Vec{n} \cdot \Vec{x}$, with $n_i = - \Lambda_i$. In order to do so, we construct the geometrical shape in $\mathbb{R}^k$ of all the points on the space of the $x_i$ that can be reached by all the different vev alignments, which we denote by $\Gamma$. Then, if $k \leq 3$ one can visualize this shape and by direct observation determine the coordinates in this shape that protrude the farthest in the direction of the vector $\Vec{n}$ for a given set of $\Lambda_i$, and from there works backwards to determine which alignments produce those $x_i$ coordinates. An example of this procedure is shown in Figure \ref{fig::gammaspace}.

\subsection{Soft-Symmetry Breaking}
Now, in order to introduce soft-symmetry breaking, we consider potentials where $v_2$ is of the general form

\begin{equation}
    v_2 = M_0 + M_i y_i.
\end{equation} We note that in this expression we continue to assume that $M_0> 0$, as well as all the $M_i$, for the sake of simplicity. However, this need not be the case, and one can consider models where some of these parameters are negative, as long as $v_2 > 0$ in some region of the $y_i$ domain.

One approach to this problem was suggested in the work where the geometric minimization was introduced \cite{Degee:2012sk}, where to proceed one would first apply a non-unitary transformation $T$ on the Higgs doublet space with the purpose of diagonalizing $v_2$, in order to bring it to the canonical form of \eqref{eq::canonical} and then follow the method as usual. This, however, proves to be technically challenging to accomplish since such a diagonalization inevitably introduces a dependence on the specific values of $M_i$ in the shape of the orbit space, making its analysis a more involved task.

We can circumvent this by relaxing the requirement that $v_2$ must be in the canonical form. If we take a look at \eqref{eq::pot_min}, we see that if $v_2$ has any $M_i \ne 0$ then the conclusion that the minimum of the potential must lie at the border of $\Gamma$ ceases to be true, since while it is still where $v_4$ attains its minimum value, one now also needs to consider the effect of $v_2$ to value of the potential. Indeed we will now find the true minimum at some other point in $\Gamma$, which can in general lie inside it.

\begin{figure}[h]
\centering
        \input{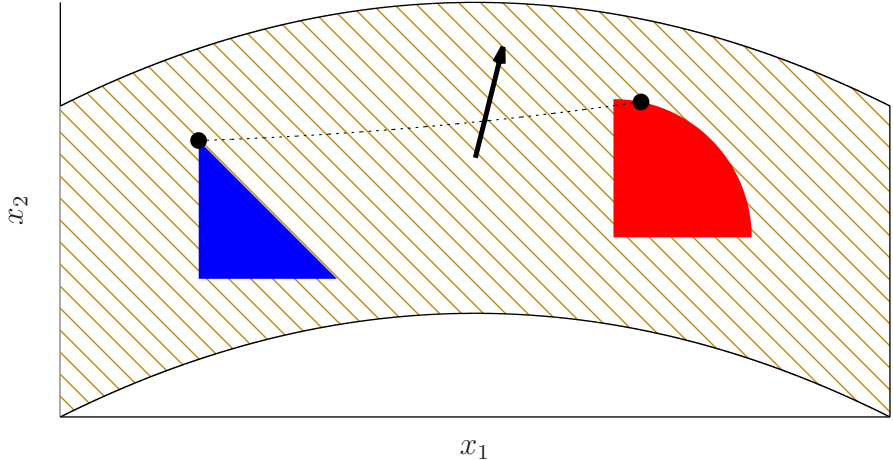}
        \caption{Examples of possible $\Xi(\alpha_i)$ obtained at different values of $\alpha_i$, shown in red and blue. The black dots represent the point that maximizes the dot product $\Vec{n} \cdot \Vec{x}$ within each $\Xi(\alpha_i)$, for the shown $\Vec{n}$, also in black. What one is interested when minimizing are the coordinates of this point as a function of $\alpha_i$. We represent an hypothetical path by the dashed line for illustrative purposes. It is important to note that this path will not be a line in the case of more than one $\alpha_i$, but will instead fill a region within $\Gamma$.}
        \label{fig::xispace}
\end{figure}

However, if we fix the value of $v_2$ by fixing the values of $y_i=\alpha_i$, with some constant set of $\alpha_i$ \footnote{We introduce the $\alpha_i$ notation to emphasize that the $y_i$ are fixed. However, they are functionally equivalent to simply using $y_i$.}, we obtain the region of $\Gamma$ compatible with that given value of $v_2$, which we denote by $\Xi(\Vec{\alpha})$ (the region depends on the set of $\alpha_i$).
Then, the minimum of this restrained problem is in the border of $\Xi(\Vec{\alpha})$, whose description now becomes a function of the $\alpha_i$. Within this region, the values of $\Lambda_i$ select the direction of steepest descent, exactly as in the case of $\Gamma$. An example is of this procedure is shown in Figure \ref{fig::xispace}.

With this, we have recast the minimization problem into the minimization of the function

\begin{equation}
\label{eq::potentialmingeneral}
    V(\Vec{\alpha}) = - \frac{v_2^2(\Vec{\alpha})}{4 v_4(\Vec{\alpha})}.
\end{equation}

\noindent From this point on, one can proceed with the usual method of minimization by derivatives, with the added detail that the domain over which the $\alpha_i$ range, henceforth denoted by $\Omega$, is bounded as one can infer from \eqref{eq::bilinear_properties}. The critical points are obtained by solving the system of equations

\begin{equation}
\label{eq::mastersystemv2}
    v_2(\Vec{\alpha})\left[2 v_4(\Vec{\alpha}) \partial_{j} v_2(\Vec{\alpha})- v_2(\Vec{\alpha}) \partial_{j} v_4(\Vec{\alpha})\right] = 0.
\end{equation}

\noindent where $\partial_{i} = \frac{\partial}{\partial {\alpha_i}}$. We can cancel the prefactor of $v_2$  by noticing that for any phenomenologically acceptable solution we must have that $v_2 > 0$ in order for the minimum to occur away from the origin leading to massive scalars. Eliminating it from \eqref{eq::mastersystemv2}, we obtain

\begin{equation}
\label{eq::mastersystem}
    2 v_4(\Vec{\alpha}) \partial_{j} v_2(\Vec{\alpha}) = v_2(\Vec{\alpha}) \partial_{j} v_4(\Vec{\alpha}).
\end{equation}
As a technical remark, when one considers more than one of the $M_i \ne 0$ it is usually easier to begin by solving

\begin{equation}
\label{eq::mastersystemsimplified}
    \partial_{i} v_2 \partial_{j} v_4 = \partial_{j} v_2 \partial_{i} v_4,
\end{equation}

\noindent due to the fact that \eqref{eq::mastersystem} still admits as a solution $v_2 = v_4 = 0$, which we know to be unphysical and is removed by \eqref{eq::mastersystemsimplified}.
\section{An Example: $S_4$ symmetric 3HDM \label{sec:example}}

We now detail the application of this method to the $S_4$ symmetric 3HDM with non-zero $(M_3, M_8)$. This symmetry group was chosen due to the fact that it has received extensive treatment with the geometric method, and thus has a large set of results we can build upon. Regarding the choice of $M_i$, it was motivated by the specific form of the potential $V_{S_4}$, leading to a simpler situation in which to illustrate the procedure.

\subsection{Fully-Symmetric Summary}
In the notation of \eqref{eq::bilinears}, the $S_4$ symmetric potential is given by \cite{Degee:2012sk}

\begin{equation}
\label{eq::potS4}
    V_{S_4} = - M_0 r_0 + \Lambda_0 r_0 ^2 + \Lambda_1 \left(r_1 ^2 + r_4 ^2 + r_6 ^2\right) + \Lambda_2 \left(r_2 ^2 + r_5 ^2 + r_7 ^2\right) + 
    \Lambda_3 \left(r_3 ^2 + r_8 ^2\right).
\end{equation}

The complete orbit space for $S_4$ has been obtained in \cite{Degee:2012sk}. Here, however, we are interested only in its neutral part. This corresponds to the dashed region that will appear later in Figure \ref{fig::orbitS4_fixed_values}.

The set of bounded from below (BFB) conditions for $S_4$ have also been obtained, even in the case of soft-symmetry breaking. In a compact form, they read \cite{Ivanov:2020jra}

\begin{align}
\label{eq::bfb}
    \Lambda_0 + \min \left(\Lambda_1, \frac{\Lambda_1}{4}, \frac{\Lambda_1 + 3\Lambda_2}{4}, \frac{\Lambda_1 + 3\Lambda_2}{16}, \frac{\Lambda_1\Lambda_2}{\Lambda_2 + 3\Lambda_1}\right) > 0,
\end{align} where for every condition one needs to add its counterpart obtained by the substitution $\Lambda_1 \rightarrow \Lambda_3$. Note that \eqref{eq::bfb} implies that all of the conditions must be met simultaneously. These will be of use in excluding phenomenologically forbidden parameter regions from the analysis to come.

\subsection{Soft-Symmetry Breaking}

As mentioned, we will focus on the case where we add non-zero $M_3$ and $M_8$, meaning that

\begin{equation}
    v_2 = M_0 + M_3 y_3 + M_8 y_8.
\end{equation}

From now on, we focus only on the neutral subspace. This simplification lets us obtain all of the possible minimum alignments as well as necessary conditions on the parameters for them to be the global minimum. If one wishes to go further and obtain necessary and sufficient conditions, two approaches can be taken: (1) either prove that $\Xi(\alpha_3, \alpha_8)$ is convex even for non-neutral vacuum configurations, at which point requiring that the squared masses of all the physical particles be positive at the neutral minimum yields the full set of sufficient and necessary conditions, or (2) in case where it is not convex conduct a full analysis of its shape.

With this remark, we proceed with the doublets parameterized as

\begin{equation}
    \label{eq::doublet_neutral}
    \Phi_i =  \begin{pmatrix}
    0 \\
    v_i e^{i\omega_i}\\ 
    \end{pmatrix}, \quad i=1, 2, 3.
\end{equation}

\subsubsection{$\Omega(y_3, y_8)$}

As discussed, $y_3$ and $y_8$ can only assume values within a certain domain $\Omega(y_3, y_8)$, which we now determine.
From their definitions and \eqref{eq::doublet_neutral} we obtain that

\begin{align}
    y_3 & = \frac{\sqrt{3}}{2}\left(v_1 ^ 2 - v_2 ^ 2 \right), \\
    y_8 & = \frac{1}{2}\left(v_1 ^ 2 + v_2 ^ 2 - 2 v_3 ^ 2\right) = \frac{1}{2}\left(1 - 3 v_3 ^ 2\right),
\end{align}

\noindent where we have set $v_1 ^ 2 + v_2 ^ 2 + v_3 ^ 2 = 1$ without loss of generality. Some straightforward manipulations then lead to 

\begin{equation}
    y_3 = \sqrt{3} \left(v_1^2 - \frac{1+y_8}{3} \right), \quad 0 \leq v_1^2 \leq \frac{2}{3}\left(1+y_8\right),
\end{equation}

\noindent which when evaluated at the minimum and maximum of $v_1^2$ yields

\begin{align}
    - \frac{1 + y_8}{\sqrt{3}} \leq & y_3 \leq \frac{1 + y_8}{\sqrt{3}}, & -1 \leq & y_8 \leq \frac{1}{2}.
\end{align}

\begin{figure}[H]
\begin{subfigure}[t]{0.45\columnwidth}
        \raisebox{-\height}{\resizebox{\columnwidth}{!}{\input{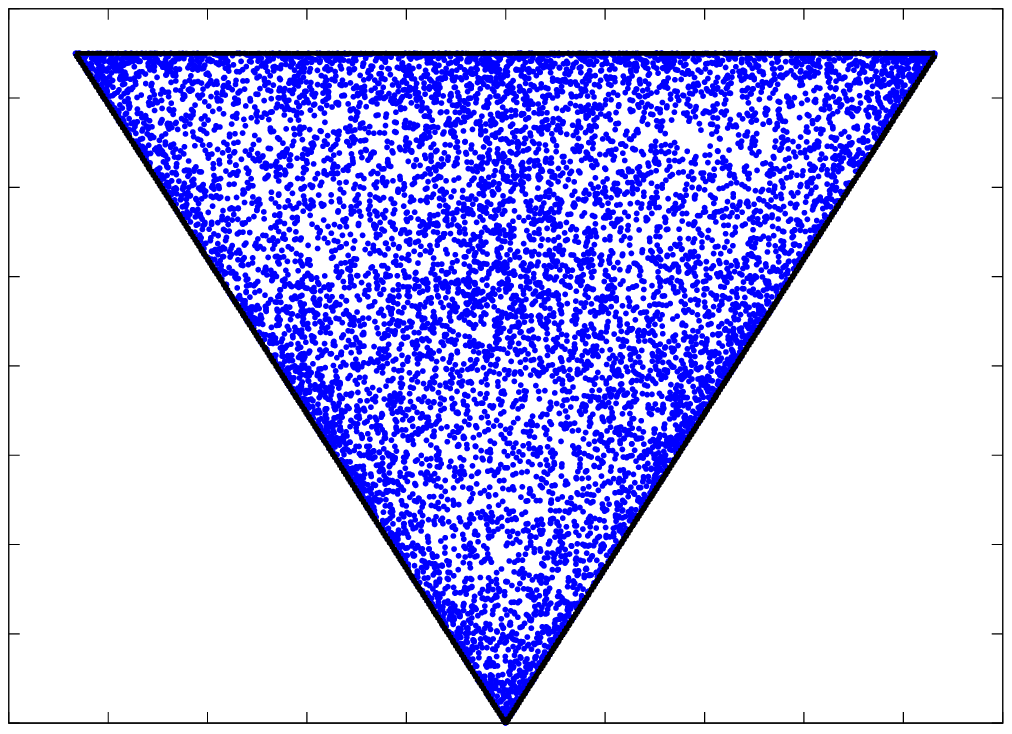}}}
        \subcaption{Complete $(y_3, y_8)$ domain. Shown in blue are the results of a numerical scan over these variables with 10000 random neutral doublet configurations.}
        \label{fig::y3y8_1}
        \end{subfigure}
        \hfill
        \begin{subfigure}[t]{0.45\columnwidth}
        \raisebox{-\height}{\resizebox{\columnwidth}{!}{\input{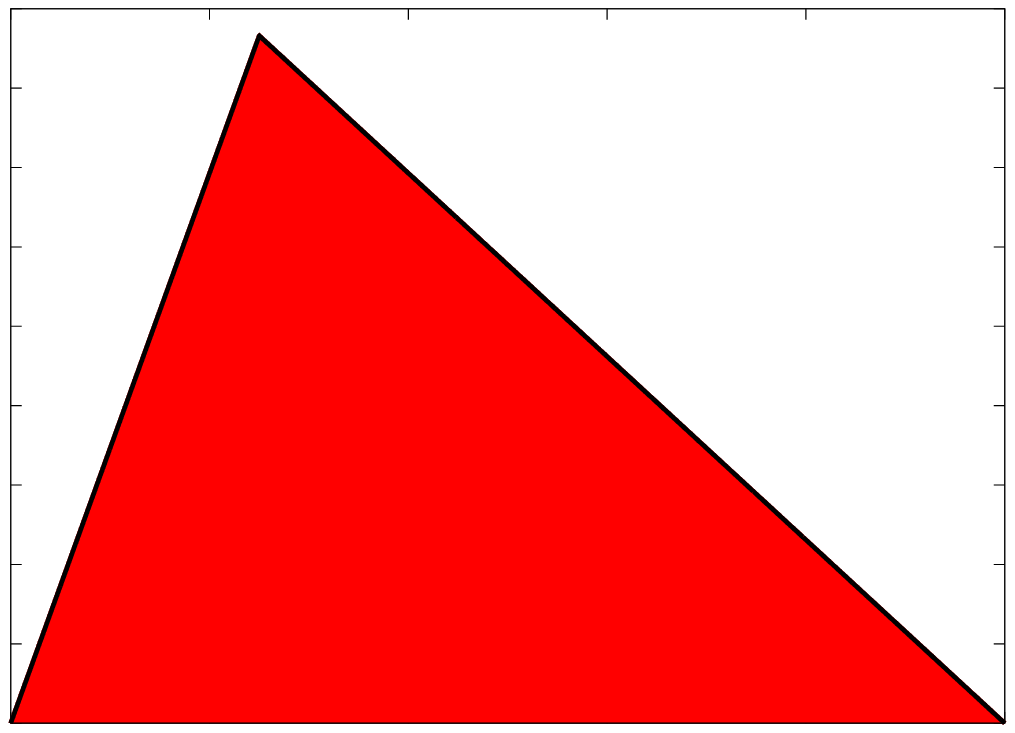}}}%
         \subcaption{Reduced domain in red obtained from symmetry considerations, expressed in terms of $({y}^\prime_3,{y}^\prime_8)$. Note how it is one sixth of the total domain, compatible with the $3!$ relative orderings of the magnitudes of the vevs.}
         \label{fig::y3y8_2}
        \end{subfigure}
          \hfill
\caption{Graphical representation of the domain}
\end{figure}

These inequalities describe an equilateral triangle with sides of length $\sqrt{3}$ and vertices at $A=(\sqrt{3}/2,1/2)$, $B=(-\sqrt{3}/2,1/2)$ and $C=(0,-1)$. We display this in Figure \ref{fig::y3y8_1}, as well as a numerical scan of the values of $y_3$ and $y_8$ obtained via a random scan over general neutral doublet configurations. We emphasize this is merely encoding the possible directions of $(v_1,v_2,v_3)$ in terms of $y_3$, $y_8$.

We start by restricting the domain of $(M_3, M_8)$. We know a priori that this is possible, due to the fact that the relative ordering of the magnitude of the doublet vevs is unphysical, and as such any specific choice is equally valid.

To see this explicitly, let us consider that $M_3 < 0$. Then, under the interchange $\Phi_1 \leftrightarrow \Phi_2$ we have that $y_3 \rightarrow -y_3$ which can be absorbed by $M_3$, making it a positive quantity. Similarly, the remaining interchanges reflect $(M_3, M_8)$ along the symmetry axes of the triangle of Figure \ref{fig::y3y8_1}. With this, we conclude that we can always perform an unphysical transformation on the doublets that brings any pair of $(M_3, M_8)$ to the region defined by 

\begin{align}\
\label{eq::domainMi}
    0 \leq & M_3 \leq \sqrt{3} M_8, & M_8 & \geq 0.
\end{align}
It is useful to introduce the dimensionless parameters

\begin{align}
    \varepsilon_3 = \frac{M_3}{\sqrt{3} M_0}, & & & \varepsilon_8 = \frac{M_8}{M_0}.
\end{align}

Furthermore, we can combine this result with the $S_3$ symmetry of the triangle in order to reduce the domain of $(y_3, y_8)$. Since the interchanges $\Phi_i \leftrightarrow \Phi_j$ are symmetries of the potential then $v_4$ is invariant under them, by definition. Thus, if we consider the specific form of \eqref{eq::pot_min} the conclusion is that for all the points connected by reflections along the axis of symmetry of Figure \ref{fig::y3y8_1} the one that leads to the lowest value of $V$ is the one that maximizes $M_i y_i$. Given the specific choice of \eqref{eq::domainMi} then the $(y_3, y_8)$ are also constrained by

\begin{align}
     0 \leq & y_3 \leq \sqrt{3} y_8, & 0 \leq y_8 & \leq \frac{1}{2}.
\end{align}

This particular choice of parameters selects  $v_1 > v_2 > v_3$. 

As a final step, it is convenient to introduce a new set of independent variables related to $(y_3, y_8)$ by a rotation of $\pi/6$ radians,

\begin{align}
\label{eq::rotationy3y8}
  &  \begin{pmatrix}
    y_3^{\prime} \\
    y_8^{\prime}
    \end{pmatrix} = \begin{pmatrix}
    \frac{\sqrt{3}}{2} & \frac{1}{2} \\
    -\frac{1}{2} & \frac{\sqrt{3}}{2}
    \end{pmatrix}  \begin{pmatrix}
    y_3 \\
    y_8
    \end{pmatrix}, & & \begin{pmatrix}
    M_3^{\prime} \\
    M_8^{\prime}
    \end{pmatrix} = \begin{pmatrix}
    \frac{\sqrt{3}}{2} & \frac{1}{2} \\
    -\frac{1}{2} & \frac{\sqrt{3}}{2}
    \end{pmatrix}  \begin{pmatrix}
    M_3 \\
    M_8
    \end{pmatrix},
\end{align}

\noindent such that the axis $y_3^{\prime}$ lies along the symmetry axis $y_3 = \sqrt{3} y_8$. We accompany this with the same rotation on the $(M_3, M_8)$ plane.

As such, the description of the relevant variables and parameters is summarized as

\begin{align}
\label{eq::ineqsparams}
    \begin{cases}
    \frac{\alpha_8^{\prime}}{\sqrt{3}} \leq \alpha_3^{\prime} \leq 1 - \sqrt{3} \alpha_8^{\prime} \\
    0 \leq \alpha_8^{\prime} \leq \frac{\sqrt{3}}{4}
    \end{cases},& & &\begin{cases}
    \varepsilon_3^{\prime} \geq \varepsilon_8^{\prime}\\
    \varepsilon_8^{\prime} \geq 0
    \end{cases},
\end{align}

\noindent with

\begin{align}
    \varepsilon_3^{\prime} = \frac{M_3^{\prime}}{M_0}, & & & \varepsilon_8^{\prime} = \frac{M_8^{\prime}}{\sqrt{3} M_0}.
\end{align} This domain is shown in Figure \ref{fig::y3y8_2}.

With these parameters, we have that

\begin{equation}
\label{eq::v2rot}
    v_2 = M_0 \left(1 + \varepsilon_3^{\prime} y_3^{\prime} + \sqrt{3} \varepsilon_8^{\prime} y_8^{\prime} \right).
\end{equation}

To avoid confusion, we note in advance that throughout Section \ref{sec:minima} we will drop the prime superscripts for the $\alpha_3^{\prime}$, $\alpha_8^{\prime}$ $\varepsilon_3^{\prime}$ and $\varepsilon_3^{\prime}$.

\subsubsection{$\Xi(\alpha_3^{\prime},\alpha_8^{\prime})$}
We now proceed to the characterization of $\Xi(\alpha_3^{\prime},\alpha_8^{\prime})$.
Given the specific form of \eqref{eq::potS4}, it is convenient to eliminate $x_2$ by using \eqref{eq::neutral_equality}, which leaves $x_1$ and $x_3$ as the quartic variables. Then,

\begin{equation}
\label{eq::v4x1x3}
    v_4 = \Lambda_{02}^+ + \Lambda_{12}^- x_1 + \Lambda_{32}^- x_3,
\end{equation} where $\Lambda_{ij}^{\pm} \equiv \Lambda_i \pm \Lambda_j$. It is important to note how this substitution reduces the number of relevant quartic free parameters from four to three, by ``absorbing'' $\Lambda_2$.
With both $\alpha_3^{\prime}$ and $\alpha_8^{\prime}$ fixed, it is straightforward to verify that

\begin{equation}
    x_3 = {\alpha_3^{\prime}}^2 + {\alpha_8^{\prime}}^2,
\end{equation}

\noindent since the primed and unprimed variables are related by a rotation. We thus conclude that $\Xi(\alpha_3^{\prime},\alpha_8^{\prime})$ will be a horizontal line in the $(x_1, x_3)$ plane. Therefore, the points of $\Xi$ that minimize $v_4$ and thus minimize the potential are the extremities of this straight line, which we denote by $L$ and $R$ for the left and right points, and the model will have two different classes of alignments.

Given the form of \eqref{eq::v4x1x3} we can also deduce under which conditions each point will correspond to the minimum: if $\Lambda_{12}^- > 0$ then the minimum of $v_4$ lies at $L$; otherwise, if $\Lambda_{12}^- < 0$ the minimum lies at $R$. This finding is in agreement with the definitions of $\Lambda_1$ and $\Lambda_2$ in \eqref{eq::potS4}, where these parameters couple to the real and imaginary parts of the mixed products of the doublets, respectively. Therefore, if for example $\Lambda_{12}^- > 0$, equivalent to $\Lambda_1 > \Lambda_2$, the configurations of doublets with non-zero relative phases lead to an overall smaller contribution to $v_4$ than their phaseless counterparts at a given magnitude, and as such lead to a smaller value of $V$.

Thus, we must determine the analytical expressions for $R$ and $L$. We have that, according to our parameterization in \eqref{eq::doublet_neutral},

\begin{align}
\label{eq::x1}
    x_1 & = \frac{3}{\left(v_1^2 + v_2^2 + v_3^2\right)^2} \left( v_1^2 v_2^2 \cos^2 \left(\delta_{23} - \delta_{13}\right) + v_1^2 v_3^2 \cos^2 \delta_{13} + v_2^2 v_3^2 \cos^2 \delta_{23}\right)\\
    & = \frac{3}{\left(q_1^2 + q_2^2 + 1\right)^2} \left(q_1^2 q_2^2 \cos^2 \left(\delta_{23} - \delta_{13}\right) + q_1^2 \cos^2 \delta_{13} + q_2^2 \cos^2 \delta_{23}\right)
\end{align} where we defined $\delta_{ij}=\omega_i - \omega_j$ and $q_1 =\frac{v_1}{v_3}$, $q_2 = \frac{v_2}{v_3}$. The introduction of the ratios $q_i$ simply reflects the fact that $x_1$ is a scale-invariant quantity and as such we are always able to eliminate one degree of freedom in the magnitude parameters $v_i$. 

Note that in general one may have alignments in which $v_3 = 0$, in which case both $q_1$ and $q_2$ will diverge. Thus, in the course of our analysis, when we encounter these divergences we need to interpret them as an alignment with null $v_3$, $(v_1, v_2, 0)$. Thus, following the point just made regarding the magnitude degree of freedom, in this case we turn our attention to the ratio $q_1/q_2 = v_1/v_2$. If we also find that this quantity diverges, we immediately know that the alignment is of the form $(v_1,0,0)$.

We start by obtaining what conditions on the $q_i$ are enforced by fixing $(y_3^\prime, y_8^\prime)=(\alpha_3^\prime, \alpha_8^\prime)$. Equating degrees of freedom, we have two $\alpha$'s and two $q$'s, and as such the latter must be fully determined by the former. Explicitly, we have that

\begin{align}
    \alpha_3^{\prime} & = \frac{1}{2} \frac{2 q_1^2 - q_2^2 - 1}{1 + q_1^2 + q_2^2}, \\
    \alpha_8^{\prime} & = \frac{\sqrt{3}}{2} \frac{q_2^2 - 1}{1 + q_1^2 + q_2^2},
\end{align}

whose solution is

\begin{align}
    q_1^2 & =\frac{1 + 2 \alpha_3^{\prime}}{1 - (\alpha_3^{\prime} + \sqrt{3} \alpha_8^{\prime})}, \\
    q_2^2 & =\frac{1 - (\alpha_3^{\prime} - \sqrt{3} \alpha_8^{\prime})}{1 - (\alpha_3^{\prime} + \sqrt{3} \alpha_8^{\prime})}.
\end{align} Combining these expression with the domain defined in \eqref{eq::ineqsparams} (which is depicted in Figure \ref{fig::y3y8_2}) and the above discussion regarding the divergences of these ratios, we can obtain an intuitive understanding of what this domain entails in terms of vev alignments: 

\begin{itemize}
    \item Inside of it these quantities are finite, and the alignment has three non-zero components $(v_1, v_2, v_3)$
    \item Along the edge $\alpha_8^{\prime} = (1 - \alpha_3^{\prime}) / \sqrt{3}$ both ratios diverge, and the alignment is of the form $(v_1, v_2, 0)$
    \item At the vertex $(\alpha_3^{\prime},\alpha_8^{\prime}) = (1, 0)$ the ratio $q_1/q_2$ also diverges, meaning that the vev alignment is of the form $(v_1, 0, 0)$.
\end{itemize} 

Given that fixing $\alpha_3^{\prime}$ and $\alpha_8^{\prime}$ fully determines $q_1$ and $q_2$, the parameters that determine the extrema of $x_1$ are the relative phases $\delta_{ij}$ of the doublets. The maximum value can be inferred by direct observation: if all $\omega_i=0$ then the cosines achieve their maximum values simultaneously and we must have the maximum of $x_1$. By substituting $q_1$, $q_2$ and null phases, we obtain that

\begin{equation}
\label{eq::x1r}
    x_1^R = 1 - \left({\alpha_3^{\prime}}^2 + {\alpha_8^{\prime}}^2\right) = 1 - x_3.
\end{equation}

This result is to be expected, since in the case of the fully-symmetric model the line $x_1 + x_3 = 1$ in the $x_i$ space corresponds to the case where all the doublets have real vevs.

Regarding $L$, the determination of its analytical expression for $x_1$ is more involved and as such is deferred to Appendix \ref{app::x1leftpoint}. The final result is the piece-wise function

\begin{equation}
\label{eq::x1l}
x_1^L = 
\begin{cases}
  \frac{1}{4} - \left({\alpha'}_3^2 + {\alpha'}_8^2\right), & 0 \leq \alpha'_3 \leq \frac{1}{4} \\
  \frac{1}{4} - \left({\alpha'}_3^2 + {\alpha'}_8^2\right) + \frac{4}{3} \left( \alpha'_3 - \frac{1}{4}\right)^2, & \frac{1}{4} < \alpha'_3 \leq 1
\end{cases},
\end{equation} with relative phases defined by

\begin{align}
0 \leq \alpha_3^{\prime} \leq \frac{1}{4}: &
\begin{cases}
  \cos{2\delta_{13}} = - \frac{1 + q_1^4 - q_2^4}{2 q_1^2} \\
  \cos{2\delta_{23}} = - \frac{1 - q_1^4 + q_2^4}{2 q_2^2} \\
  \sin{2\delta_{23}} = - \frac{q_1^2}{q_2^2} \sin{2\delta_{13}} 
\end{cases}, & \frac{1}{4} < \alpha_3^{\prime} \leq 1: \quad & \delta_{13}=\frac{\pi}{2}, \quad \delta_{23} = 0.
\end{align}

\begin{figure}[h]
\centering
        \input{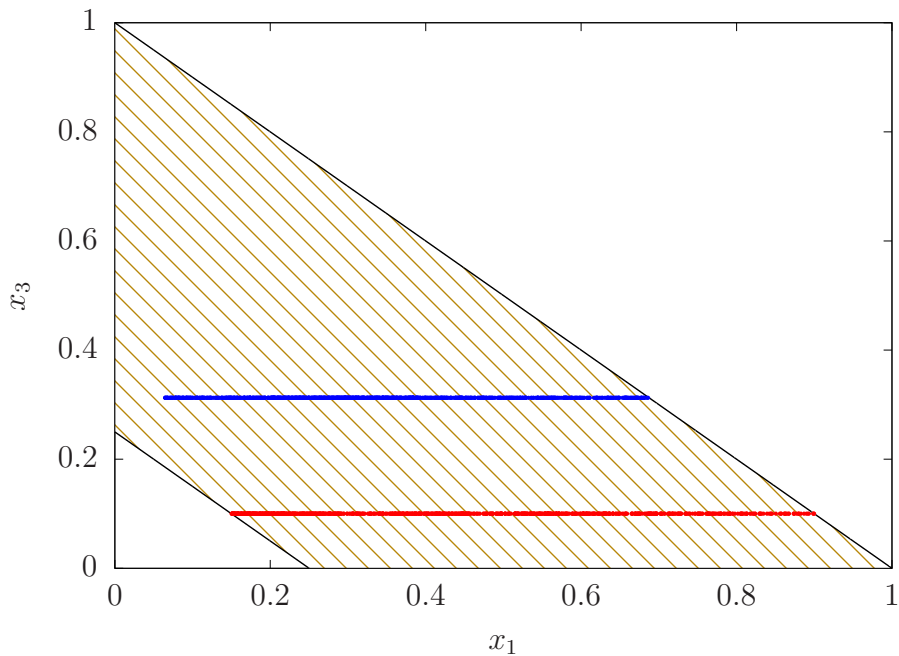}
        \caption{Neutral $\Gamma$ for $V_{S_4}$. The dashed region corresponds to the complete neutral space, filled by all neutral vev configurations, while the coloured lines exemplify two possible $\Xi(\alpha_3^{\prime}, \alpha_8^{\prime})$ regions within this space, and were obtained numerically. The red line corresponds to $(\alpha_3^{\prime},\alpha_8^{\prime}) = (0.1, 0.3)$ and the blue line to $(\alpha_3^{\prime},\alpha_8^{\prime}) = (0.5, 0.25)$. Note the agreement between the extremities of these numerical results and our analytical findings of \eqref{eq::x1r} and \eqref{eq::x1l}.}
        \label{fig::orbitS4_fixed_values}
\end{figure}

These results are shown in Figure \ref{fig::orbitS4_fixed_values} in both regimes of $\alpha_3^{\prime}$.

At this point several remarks are in order. To start, after this analysis we already have a general picture on how the different vev alignments will be like, and under what conditions on the parameters of the potential, without having fully solved the minimization problem. This insight is a direct consequence of the intuitive nature of the geometrical method, which carries over from the fully-symmetrical case.

Nevertheless, some of the shortcomings of this method must also be recognized, in order for its application to be successful. The first point to be addressed is what parameters $M_i$ one should choose to study. In our analysis we chose $(M_3, M_8)$ as the symmetry breaking parameters. However, let us suppose that we chose to consider only $M_8$ as our parameter. In that case, the value of $x_3$ would not be fixed, and one obtains for $\Xi(\alpha_8^{\prime})$ a more complex two-dimensional shape. The interested reader can obtain this shape by algebraic manipulations akin to the ones shown in Appendix \ref{app::x1leftpoint}. This is in stark contrast with the simple straight line of $\Xi(\alpha_3^{\prime},\alpha_8^{\prime})$, afforded by the specific expression of the $S_4$ potential \eqref{eq::potS4}, whose analysis then becomes a more straightforward task. Thus, in this case, if one wants to obtain the results for $M_8$ as the only non-zero parameter it is simpler to start by considering both $M_3, M_8 \ne 0$ and then set $M_3 = 0$ in the final results.

This is not to say, however, that one should adopt the strategy of adding as many parameters as possible to restrict $\Xi(\Vec{\alpha})$ to a point and then set all the undesired parameters to zero at the end of the computations. In fact, by doing so we are immediately met with algebraic complications which can stem from a number of sources. The first source is connected with the fact that we have 8 possible $y_i$ and only 4 independent parameters $q_1, q_2, \delta_{13}, \delta_{23}$, meaning that it is impossible to fix all $y_i = \alpha_i$ independently. In fact, what happens is that for more than four non-zero $M_i$ the values of $\alpha_i$ become functions of a chosen subset of 4 $\alpha_j$, whose dependence $\alpha_i(\alpha_j)$ can be non-trivial. Furthermore, even if one circumvents this, the analytical expressions obtained for the boundary of $\Xi(\Vec{\alpha})$ may themselves be non-trivial, which then makes finding the solutions of \eqref{eq::mastersystemsimplified} a highly complex task. We highlight here that the evident simplicity of the expressions for $x_1^R$ and $x_1^L$ is more of a coincidence than a general feature of this method.

Lastly, we also mention that all of the difficulties from the fully-symmetric method are evidently inherited.

The upshot from this discussion is that when applying this method one needs to keep in mind the specific structure of the potential as well as the expressions of $y_i$, and that finding solvable cases will in general be based on a trial and error approach.

\section{Minimization of $V$ \label{sec:minima}}

In this section we conduct the detailed minimization of the general form \eqref{eq::potentialmingeneral}.
We note that, as mentioned when they were introduced in Section \ref{sec:example}, in this Section we refer to the rotated $\alpha_3^{\prime}$, $\alpha_8^{\prime}$ $\varepsilon_3^{\prime}$ and $\varepsilon_3^{\prime}$ instead as $\alpha_3$, $\alpha_8$, $\varepsilon_3$, $\varepsilon_8$.

We have $v_2$ given by \eqref{eq::v2rot}. Regarding $v_4$, the results of the previous section show that two different cases must be considered, namely
\begin{equation}
     \Lambda_{12}^- < 0: v_4^R = \Lambda_{01}^+ + \Lambda_{31}^-\left({\alpha}_3^2 + {\alpha}_8^2\right),
\end{equation} and 

\begin{equation}
     \Lambda_{12}^- > 0: v_4^L = \begin{cases}
  \Lambda_{01}^+ - \frac{3}{4} \Lambda_{12}^- + \Lambda_{31}^- (\alpha_3^2 + \alpha_8^2), & 0 \leq \alpha_3 \leq \frac{1}{4}\\
  \Lambda_{01}^+ + \Lambda_{12}^- \left( -\frac{3}{4} + \frac{4}{3} (\alpha_3 - \frac{1}{4})^2 \right)+ \Lambda_{31}^- (\alpha_3^2 + \alpha_8^2), & \frac{1}{4} < \alpha_3 \leq 1
\end{cases}.
\end{equation}

The general minimization procedure is as follows. 

We start by the identification of the critical points of $V(\alpha_3, \alpha_8)$, by solving \eqref{eq::mastersystemsimplified}. Let us suppose that we have found the set of such solutions, $S$, possibly with more than one element. Then, in order for these critical points to be able to correspond to the true minimum of the potential $V$ they must simultaneously meet the validity requirements:

\begin{enumerate}
    \item[\textbf{(a)}] A positive-definite Hessian evaluated at that point
    \item[\textbf{(b)}] The point must lie inside of the domain defined by \eqref{eq::ineqsparams}
\end{enumerate}
In general, these requirements will place constraints in both the $\Lambda_i$ and the $\varepsilon_i$ to lie within a certain region of their respective spaces.

Now, if $S$ has only one element and for a given set of parameters of the potential it passes both \textbf{(a)} and \textbf{(b)} it is automatically the minimum of the potential. If, however, multiple solutions to \eqref{eq::mastersystemsimplified} satisfying \textbf{(a)} and \textbf{(b)} exist one needs to further compare them in order to determine the solution that yields the lowest value of the $V$, which will in general impose further conditions on the parameters of the potential.

However, if no element of $S$ meets both \textbf{(a)} and \textbf{(b)} we turn to the analysis of the critical points along the edges of the domain, and proceed in exactly the same manner as for the ``regular'' critical points, forming the set of solutions to \eqref{eq::mastersystemsimplified}, now denoted by $E$, to which \textbf{(a)} and \textbf{(b)} are again imposed. Note that in this case \textbf{(a)} reduces to a positive second derivative, since along the edge $V$ becomes a single variable function.

When none of the elements of $E$ are found to comply with the above conditions then we turn to the vertices of the domain. Here, we simply need to determine which vertex leads to a lower value of the potential, and under what conditions.

\subsection{$\Lambda_{12}^- < 0$}

Let us start with the simpler case of $\Lambda_{12}^- < 0$. As stated,

\begin{equation}
    v_4^R = \Lambda_{01}^+ + \Lambda_{31}^- (\alpha_3^2 + \alpha_8^2)
\end{equation} It is important to highlight that from the already reduced three parameter dependence in \eqref{eq::v4x1x3}, this expression further reduces the number of free parameters to two, completely eliminating the dependence on $\Lambda_2$. This means that in this case the analytical results for the minimum will depend only on two parameters, and the BFB conditions further demand that $\Lambda_{01}^+ > 0$. 

Applying the generic form of equation \eqref{eq::mastersystemsimplified} leads to the single solution $S_1$

\begin{align}
\label{eq::criticalpointx1R}
    \alpha_3 & = \frac{\varepsilon_3}{\overline{\Lambda}}, & \alpha_8 & = \sqrt{3} \frac{\varepsilon_8}{\overline{\Lambda}},
\end{align} 
where we have defined 

\begin{equation}
\label{eq::lambdatildeR}
\overline{\Lambda} \equiv \frac{\Lambda_{31}^-}{\Lambda_{01}^+}.
\end{equation} It turns out that this parameter solely determines the concavity of $V(\alpha_3, \alpha_8)$, so its detailed analysis is in order. As mentioned, the BFB conditions force $\Lambda_{01}^+ > 0$. Regarding $\Lambda_{31}^-$, up until now we have found no explicit requirements for it. Therefore, any possible restrictions must be imposed by the BFB conditions. In order to determine if any such restrictions exist, we write $\Lambda_3$ as

\begin{equation}
    \Lambda_3 = \Lambda_1 +  \Lambda_{01}^+ \overline{\Lambda}.
\end{equation} Substituting into \eqref{eq::bfb} we find the particularly interesting condition

\begin{equation}
    \Lambda_0 + \Lambda_3 > 0 \Rightarrow \Lambda_{01}^+(1 + \overline{\Lambda}) > 0.
\end{equation} Since $\Lambda_{01}^+ > 0$, we must have that $\overline{\Lambda} > -1$, meaning that the BFB conditions imply that this parameter is also (coincidentally) bounded from below. Regarding the remaining BFBs, a simple graphical analysis on the $(\Lambda_0, \Lambda_1)$ plane shows that whatever the value of $\overline{\Lambda} > -1$ that one chooses, it is always possible to find values of $\Lambda_0$ and $\Lambda_1$ such that all BFB are met at the same time. As such, we conclude that there exists this single lower limit on $\overline{\Lambda}$.
Following along the general procedure, imposing \textbf{(a)} and \textbf{(b)} simultaneously results in

\begin{equation}
    \overline{\Lambda} > 0.
\end{equation}
This implies that once again we are met with two different regimes, separated by the sign of $\overline{\Lambda}$.

\subsubsection{$\overline{\Lambda} > 0$}

In this case, $S_1$ (from (\ref{eq::criticalpointx1R})) corresponds to the minimum, provided that

\begin{align}
   \varepsilon_3 &\leq \overline{\Lambda}, & \varepsilon_8 & \leq \frac{\overline{\Lambda} - \varepsilon_3}{3}.
\end{align} These conditions ensure that $S_1$ lies inside the allowed domain for $\alpha_3$ and $\alpha_8$.
If they are not met we turn to the edges of the domain. However, in this specific case we can forego the full analysis of its three sides by noticing that when we break these conditions the critical point of an unrestricted version of $V(\alpha_3,\alpha_8)$ lies the closest to the edge defined by $\alpha_8 = (1 - \alpha_3)/\sqrt{3}$, and therefore this edge must lead to the lowest value of $V$ at its critical point. Denoting this point by $E_1$, we obtain by solving \eqref{eq::mastersystemsimplified} that its explicit expression is

\begin{align}
\alpha_3 & = \frac{\overline{\Lambda}(1 + \varepsilon_3) + 3 (\varepsilon_3 - \varepsilon_8)}{\overline{\Lambda}\left[4 \left(1 + \varepsilon_3\right) - 3 \left( \varepsilon_3 - \varepsilon_8\right)\right]}, \\
\alpha_8 & = \sqrt{3} \frac{\overline{\Lambda}(1 + \varepsilon_8) - (\varepsilon_3  - \varepsilon_8)}{\overline{\Lambda}\left[4(1 + \varepsilon_3) - 3 (\varepsilon_3 - \varepsilon_8)\right]}.
\end{align}
For this point to be inside the allowed domain we need $\frac{1}{4} \leq \alpha_3 \leq 1$. From its expression and the constraints in \eqref{eq::ineqsparams} it is clear that the left inequality is always satisfied. Regarding the right inequality, it is met when

\begin{equation}
    \varepsilon_8 \geq \frac{\varepsilon_3 - \overline{\Lambda}}{1 + \overline{\Lambda}}.
\end{equation}
If the conditions on $\varepsilon_8$ are not satisfied, then the minimum lies at the point $(\alpha_3,\alpha_8)=(1,0)$.

\subsubsection{$-1 < \overline{\Lambda} < 0$}
Contrary to the previous case, $S_1$ is discarded. A simple analysis of the three edges of the domain leads to the same conclusion. As such, it follows that the minimum is at a vertex. Comparing the values of $V$ at the three vertices it is straightforward to verify that the point $(\alpha_3, \alpha_8)=(1,0)$ always leads to the smallest value of the potential, and is therefore the minimum. Thus, in this case we are presented with a rigid structure for the vev alignment, regardless of the specific values of the parameters.

\subsubsection{The role of $\overline{\Lambda}$}

A remarkable feature of the results just shown is their exclusive dependence on $\overline{\Lambda}$, in the sense that this is the only combination of the quartic parameters that appears in the expressions and inequalities we have derived. In this sense, this parameter fully encapsulates the influence of the quartic coupling parameters in the selection of the minimum alignment, effectively mapping the initial two dimensional dependence (in $\Lambda_{01}^+$ and $\Lambda_{31}^+$) to a single dimension \footnote{Note that the dependence on $\Lambda_{01}^+$ has not been eliminated. In fact, it will influence the magnitude of the alignment, $r_0$.}. Furthermore, it greatly facilitates one's understanding of what the BFB conditions entail in terms of allowed parameter ranges, by allowing a factorization of the four quartic parameters into $1+3$, leading to a more tractable geometric treatment. Lastly, it also acts as a sort of discriminant for the model, since its sign separates two different alignment regimes.

\subsection{$\Lambda_{12}^- > 0$}
Here, our starting point is
\begin{equation}
v_4^L = \begin{cases}
  \Lambda_{01}^+ - \frac{3}{4} \Lambda_{12}^- + \Lambda_{31}^- (\alpha_3^2 + \alpha_8^2) ,& 0 \leq \alpha_3 \leq \frac{1}{4}\\
  \Lambda_{01}^+ + \Lambda_{12}^- \left( -\frac{3}{4} + \frac{4}{3} (\alpha_3 - \frac{1}{4})^2 \right)+ \Lambda_{31}^- (\alpha_3^2 + \alpha_8^2), & \frac{1}{4} < \alpha_3 \leq 1
\end{cases},
\end{equation} where $\Lambda_{01}^+> 0$, $\Lambda_{12}^- > 0$ and $ \Lambda_{31}^-$ can assume positive or negative values. 
Computing the solutions of \eqref{eq::mastersystemsimplified}, we arrive at the single solution $S_1$

\begin{align}
    \label{eq::s1lt14}
        \alpha_3 & = \frac{\varepsilon_3}{4\overline{\Lambda}_1}, & \alpha_8 & = \sqrt{3} \frac{\varepsilon_8}{4\overline{\Lambda}_1}.
    \end{align} for $0 \leq \alpha_3 \leq \frac{1}{4}$, and
    
\begin{align}
\label{eq::s1gt14a3}
        \alpha_3 & = \frac{(\overline{\Lambda}_2 - 4 \overline{\Lambda}_1)(1 + \varepsilon_3) + 3\varepsilon_3(1 + \overline{\Lambda}_1)}{(\overline{\Lambda}_2 - 4 \overline{\Lambda}_1)(1 + \varepsilon_3) + 3\overline{\Lambda}_2(1 + \overline{\Lambda}_1)}, \\
        \alpha_8 & = \frac{\sqrt{3} \varepsilon_8 (1 + \overline{\Lambda}_1)(\overline{\Lambda}_2 - \overline{\Lambda}_1)}{\overline{\Lambda}_1(\overline{\Lambda}_2 - 4 \overline{\Lambda}_1)(1 + \varepsilon_3) + 3\overline{\Lambda}_2(1 + \overline{\Lambda}_1)}.
        \label{eq::s1gt14a8}
\end{align} for $\frac{1}{4} < \alpha_3 \leq 1$. In these expressions we have defined

\begin{align}
\label{eq::l1l2}
    \overline{\Lambda}_1 & \equiv \frac{\Lambda_{31}^-}{4\Lambda_{01}^+ - 3\Lambda_{12}^-}, & \overline{\Lambda}_2 & \equiv \frac{\Lambda_{12}^- + \Lambda_{31}^-}{\Lambda_{01}^+ - \Lambda_{12}^-}.
\end{align} As we will see these parameters play the same role as $\overline{\Lambda}$ on the previous analysis. This particular choice's motivation was that it allows for a simple identification of the valid ranges of $(\varepsilon_3, \varepsilon_8)$. For example, from \eqref{eq::s1lt14} it is evident that $\varepsilon_3 \leq \overline{\Lambda}_1$ for the critical point to lie inside the domain.
With the adequate parameters defined, the next step is the determination of the restrictions imposed on them both by $\Lambda_{12}^- > 0$ and by the BFB conditions in \eqref{eq::bfb}. Inverting the relations \eqref{eq::l1l2} we obtain that

\begin{align}
    \Lambda_{12}^- & = \Lambda_{01}^+ \frac{\overline{\Lambda}_2 - 4 \overline{\Lambda}_1}{1 + \overline{\Lambda}_2 - 3 \overline{\Lambda}_1}, & \Lambda_{31}^- & = \Lambda_{01}^+ \frac{ \overline{\Lambda}_1(4 + \overline{\Lambda}_2) }{1 + \overline{\Lambda}_2 - 3 \overline{\Lambda}_1}.
\end{align} Requiring positivity of $\Lambda_{12}^-$ and taking into account that the BFB conditions imply that $\Lambda_{01}^+ > 0$ we encounter two possibilities:

\begin{align}
\label{eq::l1l2regions}
   \mathbf{I}:& \begin{cases}
      \overline{\Lambda}_2 - 4 \overline{\Lambda}_1 > 0 \\
      1 + \overline{\Lambda}_2 - 3 \overline{\Lambda}_1 > 0
    \end{cases},  & \mathbf{II}:& \begin{cases}
      \overline{\Lambda}_2 - 4 \overline{\Lambda}_1 < 0 \\
      1 + \overline{\Lambda}_2 - 3 \overline{\Lambda}_1 < 0
    \end{cases}.
\end{align}
Now, when we substitute $\Lambda_2$ and $\Lambda_3$ as functions of $\overline{\Lambda}_1$ and $\overline{\Lambda}_2$ into the full set of BFB conditions, three of them involve solely $\overline{\Lambda}_1$, $\overline{\Lambda}_2$ and $\Lambda_{01}^+$, namely

\begin{align}
    \Lambda_0  + \frac{\Lambda_1 + 3 \Lambda_2}{4} & > 0 \Rightarrow \Lambda_{01}^+ \frac{4 + \overline{\Lambda}_2}{4(1 - 3 \overline{\Lambda}_1 + \overline{\Lambda}_2)} > 0 \label{eq::bfb1},\\
    \Lambda_0 + \frac{\Lambda_3 + 3 \Lambda_2}{4} & > 0 \Rightarrow \Lambda_{01}^+ \frac{(1 + \overline{\Lambda}_1)(4 + \overline{\Lambda}_2)}{4(1 - 3 \overline{\Lambda}_1 + \overline{\Lambda}_2)} > 0 \label{eq::bfb4}, \\
    \Lambda_0 + \Lambda_3 & > 0 \Rightarrow \Lambda_{01}^+ \left( 1 + \frac{\overline{\Lambda}_1 (4 + \overline{\Lambda}_2)}{1 - 3 \overline{\Lambda}_1 + \overline{\Lambda}_2} \right)> 0 \label{eq::bfb7}.
\end{align} Combining \eqref{eq::bfb1} and \eqref{eq::bfb4} we can immediately conclude that $\overline{\Lambda}_1 > -1$. Note here the identical lower bound to $\overline{\Lambda}$ of the previous section, despite their different definitions.
Regarding the two other conditions we must separate their analysis according to the two possible cases identified for $\Lambda_{12}^- > 0$.

\begin{figure}[h]
\centering
        \input{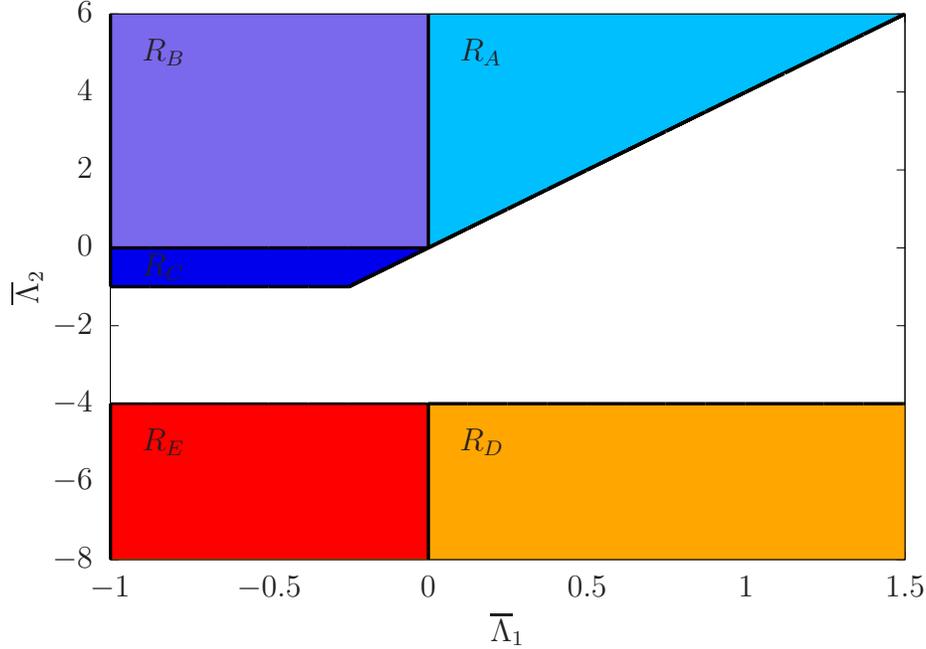}
        \caption{Partition of the $(\overline{\Lambda}_1, \overline{\Lambda}_2)$ plane. The two disjoint regions correspond to the conditions imposed by the BFB conditions. Then, within these there are further separations, shown with different tones. These correspond to regions where of different validity of the critical points of $V$.}
        \label{fig::l1l2criticalpoints}
\end{figure}

\begin{enumerate}
    \item[\textbf{I}:] In this case, from \eqref{eq::bfb1} comes that $\overline{\Lambda}_2 > - 4$, since its denominator is positive by hypothesis. Furthermore, this also implies that \eqref{eq::bfb7} reduces to $(1 + \overline{\Lambda}_1)(1 + \overline{\Lambda}_2) > 0$, imposing $\overline{\Lambda}_2 > -1$ which is a stronger condition than $\overline{\Lambda}_2 > -4$, and as such takes precedence over it.
    
    \item[\textbf{II}:] Here, the negative denominator of \eqref{eq::bfb1} implies that $\overline{\Lambda}_2 < -4$. Regarding \eqref{eq::bfb7}, it now reads $(1 + \overline{\Lambda}_1)(1 + \overline{\Lambda}_2) < 0$, and thus is satisfied when $\overline{\Lambda}_2 < -1$. Contrary to \textbf{I} this condition is now the least restraining, meaning that we take $\overline{\Lambda}_2 < -4$.
\end{enumerate}
Summarizing the results one is then led to the two disjoint regions depicted in Figure \ref{fig::l1l2criticalpoints}.
Regarding the remaining 7 BFB conditions, a similar graphical analysis to the case of $\Lambda_{12}^- < 0$ in the $(\Lambda_0, \Lambda_1)$ plane shows that they do not further constrain the $\overline{\Lambda}$ - with their values already obeying the BFB as outlined above, one is always able to find values of $\Lambda_0$ and $\Lambda_1$ such that all BFB are simultaneously met.
Now, imposing \textbf{(a)} and \textbf{(b)} on top of the results for $\overline{\Lambda}_1$ and $\overline{\Lambda}_2$, we obtain the single restriction for $S_1$ in \eqref{eq::s1lt14} and \eqref{eq::s1gt14a3}, \eqref{eq::s1gt14a8}

\begin{equation}
    \overline{\Lambda}_1 > 0,
\end{equation} on the quartic parameters.

At this point, we could proceed as in the previous section, considering two different cases according to the sign of $\overline{\Lambda}_1$. However, the exposition of results becomes clearer if we instead further subdivide the $(\overline{\Lambda}_1,\overline{\Lambda}_2)$ plane according to the validity of the critical points at the edges. A simple analysis leads to the regions shown in Figure \ref{fig::l1l2criticalpoints}. Within these, the following critical points are valid:

\begin{enumerate}
    \item[$\mathbf{R_A}$:] This region is defined by $ \overline{\Lambda}_2 > 0, \; 0 < \overline{\Lambda}_1 \leq \frac{\overline{\Lambda}_2}{4}$. Here, $S_1$ is valid, for both \eqref{eq::s1lt14} and \eqref{eq::s1gt14a3}, \eqref{eq::s1gt14a3}. Furthermore, all critical points along the edges are valid as well. 
    \item[$\mathbf{R_B}$:] This region is defined by $\overline{\Lambda}_2 > 0, \; -1 < \overline{\Lambda}_1 \leq 0$. Only the critical points along the edges $\alpha_8 = (1- \alpha_3)/\sqrt{3}$ and $\alpha_8 = 0$,$\alpha_3>1/4$ are valid.
    \item[$\mathbf{R_C}$:] This region is defined by $\overline{\Lambda}_2 < 0, \; -1 < \overline{\Lambda}_1 \leq \frac{\overline{\Lambda}_2}{4}$. No critical point is valid.
    \item[$\mathbf{R_D}$:] This region is defined by $\overline{\Lambda}_2 < -4, \; 0 < \overline{\Lambda}_1$, and has the same critical points as $R_A$.
    \item[$\mathbf{R_E}$:] This region is defined by $\overline{\Lambda}_2 < -4, \; -1 < \overline{\Lambda}_1 < 0$, and has the same critical points as $R_B$.
\end{enumerate}

With this partition of the $(\overline{\Lambda}_1,\overline{\Lambda}_2)$ parameter space it is clear that we now have five different cases to analyze. We simply present the results without proof, since obtaining them is a simple matter of imposing conditions \textbf{(a)} and \textbf{(b)} and the manipulation of inequalities, which would clutter the exposition of results. We do however note that in the treatment of $S_1$ (of \eqref{eq::s1gt14a3} and \eqref{eq::s1gt14a8}) it is easier to analyze the properties of the Hessian if one studies its trace and determinant rather than the eigenvalues directly, since the former are significantly simpler than the latter. Before we proceed, it is convenient to introduce the analytic expression for the critical point along the edge defined by $\alpha_8 = ( 1 - \alpha_3 ) / \sqrt{3}$:

\begin{align}
\label{eq::minedgea3}
        \alpha_3 & = \frac{\overline{\Lambda}_2(1 + \varepsilon_3) + 3 (\varepsilon_3  - \varepsilon_8)}{\overline{\Lambda}_2\left[4(1 + \varepsilon_3) - 3 (\varepsilon_3 - \varepsilon_8)\right]}, \\
          \alpha_8 & = \sqrt{3} \frac{\overline{\Lambda}_2(1 + \varepsilon_8) - (\varepsilon_3  - \varepsilon_8)}{\overline{\Lambda}_2\left[4(1 + \varepsilon_3) - 3 (\varepsilon_3 - \varepsilon_8)\right]}
          \label{eq::minedgea8}
\end{align}

\subsubsection{$\mathbf{R_A}:$}
The critical point $S_1$ in both forms \eqref{eq::s1lt14} and \eqref{eq::s1gt14a3}, \eqref{eq::s1gt14a8} is valid if

\begin{align}
    0 \leq \alpha_3 \leq \frac{1}{4}: \qquad & \varepsilon_3 \leq \overline{\Lambda}_1. \\
    \frac{1}{4} < \alpha_3 \leq 1: \qquad & \overline{\Lambda}_1 < \varepsilon_3 \leq \overline{\Lambda}_2, \qquad \varepsilon_8 \leq \frac{\overline{\Lambda}_1}{\overline{\Lambda}_2 - \overline{\Lambda}_1} \left(\overline{\Lambda}_2 - \varepsilon_3 \right).
\end{align} If the SBPs are such that these conditions are not met, then the minimum occurs at one of the edges. Comparing the values of the potential at each of these one obtains that the deepest one occurs at the one defined by $\alpha_8 = ( 1 - \alpha_3 ) / \sqrt{3}$, valid if

\begin{equation}
    \varepsilon_8 \geq \frac{\varepsilon_3 - \overline{\Lambda}_2}{1 + \overline{\Lambda}_2}.
\end{equation} Otherwise, the vertex $(\alpha_3, \alpha_8)=(1, 0)$ is the minimum of the potential.

\subsubsection{$\mathbf{R_B}:$}

Within this region, both forms of $S_1$ \eqref{eq::s1lt14} and \eqref{eq::s1gt14a3},\eqref{eq::s1gt14a8} cease to be valid, and we turn to the critical points $E$. Of those, the edge $\alpha_8 = (1 - \alpha_3)/\sqrt{3}$ always leads to the lowest value of the potential. It is valid if, as before,

\begin{equation}
      \varepsilon_8 \geq \frac{\varepsilon_3 - \overline{\Lambda}_2}{1 + \overline{\Lambda}_2}.
\end{equation} Otherwise, the minimum occurs at the vertex $(\alpha_3, \alpha_8)=(1, 0)$.

\subsubsection{$\mathbf{R_C}$}

In this case, since neither $S$ nor $E$ have valid elements we conclude that the minimum must lie at a vertex. Comparing the value of the potential at the three points, the conclusion is that $(\alpha_3, \alpha_8)=(1,0)$ produces the minimum alignment.

\subsubsection{$\mathbf{R_D}$}
Here, $S_1$ of \eqref{eq::s1lt14} and \eqref{eq::s1gt14a3},\eqref{eq::s1gt14a8} is valid, under the conditions

\begin{align}
    0 \leq \alpha_3 \leq \frac{1}{4}: \qquad & \varepsilon_3 \leq \overline{\Lambda}_1. \\
    \frac{1}{4} < \alpha_3 \leq 1: \qquad & \overline{\Lambda}_1 < \varepsilon_3, \qquad \varepsilon_8 \leq \frac{\overline{\Lambda}_1}{\overline{\Lambda}_2 - \overline{\Lambda}_1} \left(\overline{\Lambda}_2 - \varepsilon_3 \right).
\end{align} Note that these are almost identical to the situation with positive $\overline{\Lambda}_2$, differing only in the missing upper bound on $\varepsilon_3$. If these conditions are not met, then the minimum is at the point given by \eqref{eq::minedgea3}, \eqref{eq::minedgea8}.

\subsubsection{$\mathbf{R_E}$}
Finally, within this region we have that only $E$ has valid critical points, and of those the one of \eqref{eq::minedgea3}, \eqref{eq::minedgea8} always yields the global minimum alignment, without further restrictions on $\varepsilon_3$ and $\varepsilon_8$.

\subsubsection{Summary of Results}

Having determined all the possible minima and the conditions under which they occur, in this section we compile all the above inequalities on the SBPs into graphical representations on the $(\varepsilon_3,\varepsilon_8)$ plane, in which we also overlay the results of a numerical minimization of \eqref{eq::potentialmingeneral}, with the common parameters $M_0 = 5$, $\Lambda_{01}^+ = 4$ and $0 \leq \varepsilon_3 < 0.15$, $0 \leq \varepsilon_8 < 0.1$. These ranges were simply chosen in order to obtain representations in which all features are clearly visible.
Regarding the colouring, blue corresponds to $S_1$; in the case $\Lambda_{12}^-> 0$ where $S_1$ has two separate analytical expressions we differentiate between them by assigning the darker tone to $\alpha_3 \leq 1/4$, where it is given by \eqref{eq::s1lt14}, and the lighter to $1/4 < \alpha_3 < 1$, where its expression is \eqref{eq::s1gt14a3}, \eqref{eq::s1gt14a8}; red corresponds to $E_1$ and orange to the vertex $(1,0)$.

\begin{figure}[h]
\begin{subfigure}[t]{0.45\columnwidth}
        \raisebox{-\height}{\resizebox{\columnwidth}{!}{\input{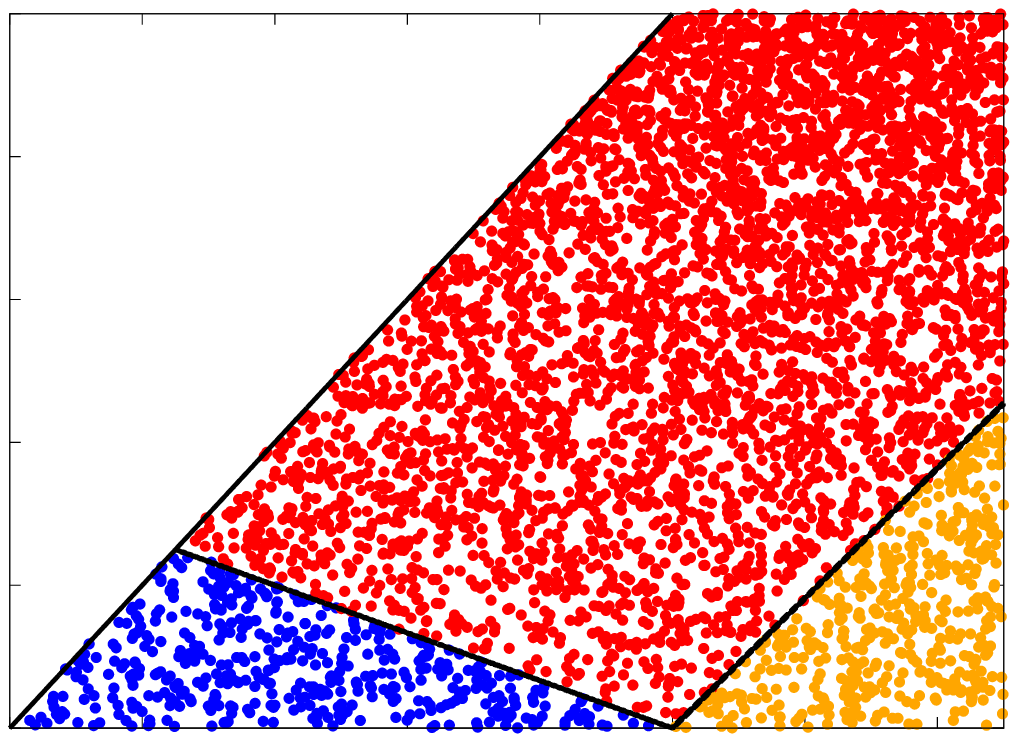}}}
        \subcaption{$\overline{\Lambda} = 0.1$}
        \end{subfigure}
        \hfill
        \begin{subfigure}[t]{0.45\columnwidth}
        \raisebox{-\height}{\resizebox{\columnwidth}{!}{\input{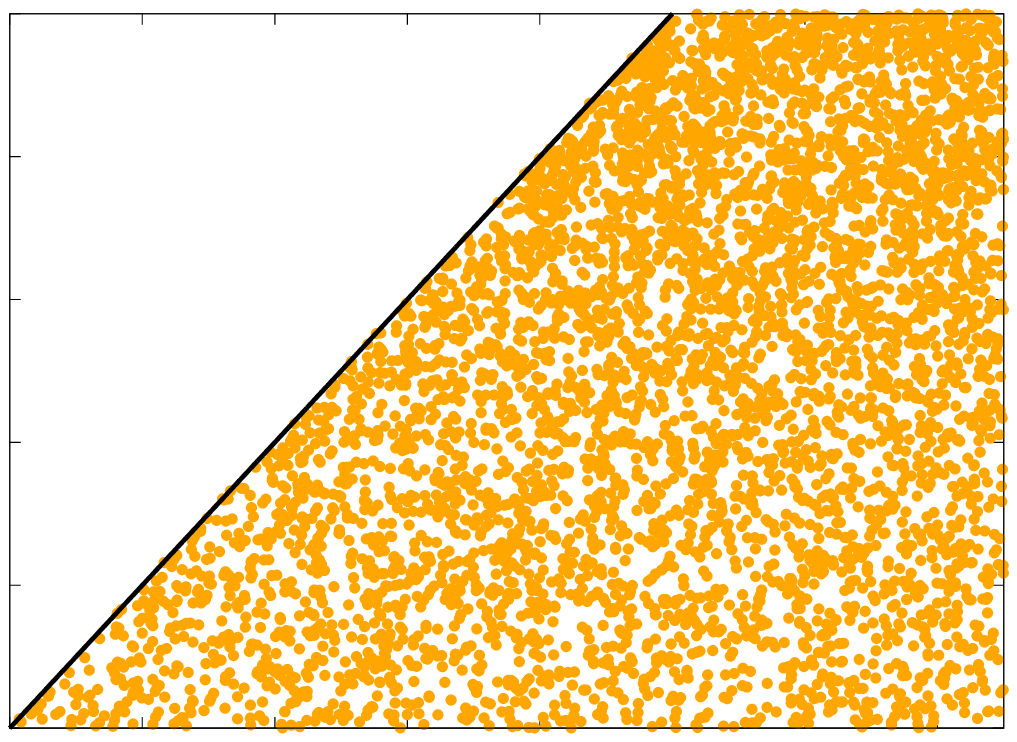}}}%
         \subcaption{$\overline{\Lambda} = -0.1$}
        \end{subfigure}
        \caption{Types of minima for $\Lambda_{12}^- < 0$.}
        \label{fig::right}
\end{figure}

\begin{figure}[H]
        \begin{subfigure}[t]{0.45\columnwidth}
        \raisebox{-\height}{\resizebox{\columnwidth}{!}{\input{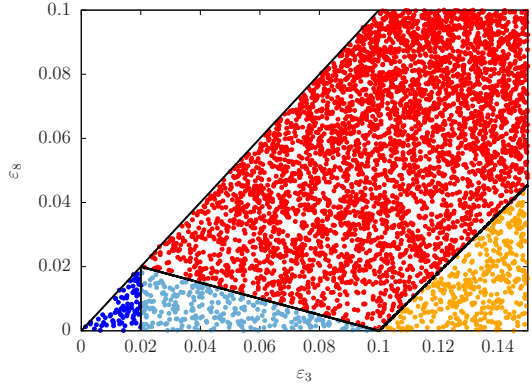}}}%
         \subcaption{$\mathbf{R_A}: \; \overline{\Lambda}_1 = 0.02$,\; $\overline{\Lambda}_2 = 0.1$}
        \end{subfigure}
          \hfill
        \begin{subfigure}[t]{0.45\columnwidth}
        \raisebox{-\height}{\resizebox{\columnwidth}{!}{\input{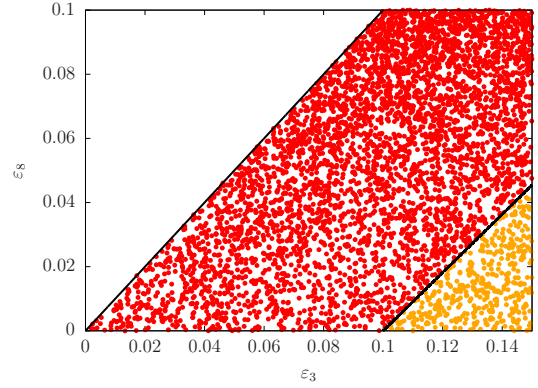}}}%
         \subcaption{$\mathbf{R_B}: \;\overline{\Lambda}_1 = -0.02$,\; $\overline{\Lambda}_2 = 0.1$}
        \end{subfigure}
        \hfill
        \begin{subfigure}[t]{0.45\columnwidth}
        \raisebox{-\height}{\resizebox{\columnwidth}{!}{\input{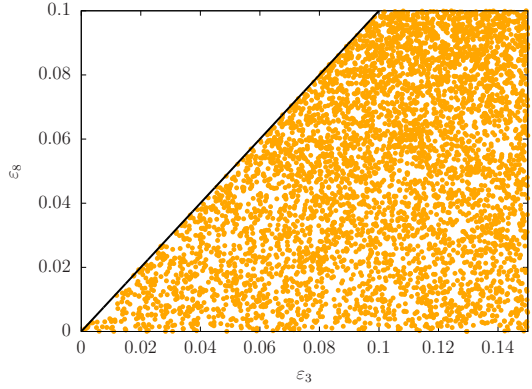}}}
        \subcaption{$\mathbf{R_C}: \;\overline{\Lambda}_1 = -0.02$,\; $\overline{\Lambda}_2 = -0.1$}
        \end{subfigure}
        \hfill
        \begin{subfigure}[t]{0.45\columnwidth}
        \raisebox{-\height}{\resizebox{\columnwidth}{!}{\input{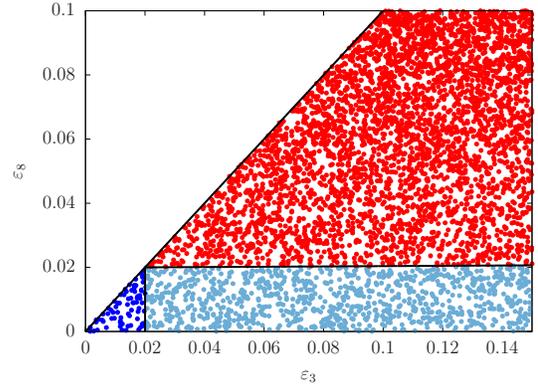}}}
        \subcaption{$\mathbf{R_D}: \;\overline{\Lambda}_1 = 0.02$,\; $\overline{\Lambda}_2 = -4.1$}
        \end{subfigure}
        \hfill
        \begin{subfigure}[t]{0.45\columnwidth}
        \raisebox{-\height}{\resizebox{\columnwidth}{!}{\input{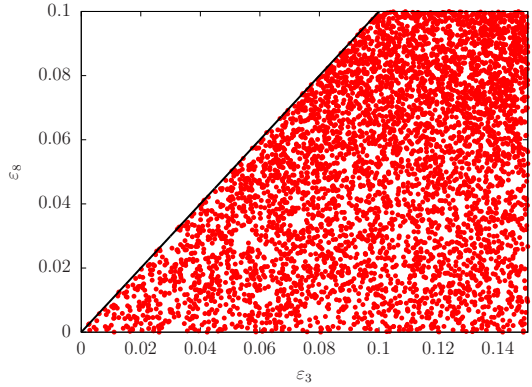}}}
        \subcaption{$\mathbf{R_E}: \;\overline{\Lambda}_1 = -0.02$,\; $\overline{\Lambda}_2 = -4.1$}
        \end{subfigure}
        \caption{Types of minima for $\Lambda_{12}^- > 0$.}
        \label{fig::left}
\end{figure}

We depict the case of $\Lambda_{12}^- < 0$ in the diagrams of Figure \ref{fig::right} and the case of $\Lambda_{12}^-> 0$ in Figure \ref{fig::left}.

From these results we highlight the similarities between both signs of $\Lambda_{12}^-$, afforded by the specific definitions of the auxiliary parameters $\overline{\Lambda}$, $\overline{\Lambda}_1$ and $\overline{\Lambda}_2$.

\section{Conclusion \label{sec:con}}

We have extended the method of Geometric Minimization to softly-broken Multi-Higgs Doublet Model potentials. This powerful method is very appropriate to potentials invariant under large symmetries, which contain few parameters, but it works well with the soft-breaking parameters and allows the classification of the possible vacuum expectation values.

The new methodology contrasts to the existing proposal as it does not consist in employing a unitary transformation that would bring back the softly-broken quadratic terms to canonical form, but in doing so, introduce complications in the quartic terms. Instead, we propose to perform a constrained geometric minimization by keeping the symmetric quartic terms while considering the effect that the softly-broken quadratic terms have in the alignment of the minimum.

We exemplify the new method in softly-broken $S_4$ invariant potential with specific soft-breaking parameters. In the following analysis, we show the relevance of characterizing the allowed domains, and systematically present the minimum for each relevant region of parameter space, as divided by analytical inequalities that apply to the quartic couplings. As we note, the conditions obtained from requiring the potential to be bounded from below powerfully restrict the quartics. We stress that, in our examples, a careful definition of the relevant combinations of quartics allows to greatly simplify the respective inequalities into two main regions of parameter space, one which splits into two classes and the other into five classes of behaviour in terms of the possible minima.

\section*{Acknowledgments}
IdMV acknowledges funding from Funda\c{c}\~{a}o para a Ci\^{e}ncia e a Tecnologia (FCT) through the contract UID/FIS/00777/2020 and was supported in part by FCT through projects CFTP-FCT Unit 777 (UID/FIS/00777/2019), PTDC/FIS-PAR/29436/2017, CERN/FIS-PAR/0004/2019 and CERN/FIS-PAR/0008/2019 which are partially funded through POCTI (FEDER), COMPETE, QREN and EU.
DI acknowledges funding from Funda\c{c}\~{a}o para a Ci\^{e}ncia e a Tecnologia (FCT) through project CERN/FIS-PAR/0008/2019 and NuPhys - CERN/FIS-PAR/0004/2019.

\appendix

\section{Minimization of $x_1$}
\label{app::x1leftpoint}

In this Appendix we detail the determination of the left extremity of $x_1$. The first step is the linearization of \eqref{eq::x1} by means of the cosine double angle identity, leading to

\begin{align}
\label{eq::x1leftapp}
    x_1 & = \frac{3}{\left(1 + q_1^2 + q_2^2\right)^2} \left(q_1^2 q_2^2 \cos^2 \left(\delta_{23} - \delta_{13}\right) + q_1^2 \cos^2 \delta_{13} + q_2^2 \cos^2 \delta_{23}\right) \\
    & = x_1^0 + x_1^\delta,
\end{align} where 

\begin{align}
    x_1^0 = & \frac{3}{2\left(1 + q_1^2 + q_2^2\right)^2}
    \left(q_1^2 q_2^2 + q_1^2 + q_2^2 \right), \\ 
    x_1^\delta = & \frac{3}{2\left(1 + q_1^2 + q_2^2\right)^2}
    \left(q_1^2 q_2^2 \cos 2\left(\delta_{23} - \delta_{13}\right) + q_1^2 \cos 2\delta_{13} + q_2^2 \cos 2\delta_{23}\right).
\end{align} Differentiating $x_1^\delta$ w.r.t $\delta_{13}$ and $\delta_{23}$ and equating to zero, one obtains

\begin{align}\label{eq::diffd13}
   \frac{\partial x_1^\delta}{\partial \delta_{13}}:\qquad 0 & = q_1^2 \left( -\sin 2\delta_{13} + q_2^2 \sin 2 (\delta_{23} - \delta_{13}) \right), \\
    \label{eq::diffd23}
   \frac{\partial x_1^\delta}{\partial \delta_{23}}:\qquad  0 & = q_2^2 \left( \sin 2\delta_{23} + q_1^2 \sin 2 (\delta_{23} - \delta_{13}) \right),
\end{align} where we know from the main text that $q_1 \ne 0$. Subtracting \eqref{eq::diffd13} from \eqref{eq::diffd23} leads to

\begin{equation}
\label{eq::relationsines}
    0 = q_1^2 \sin 2 \delta_{13} + q_2^2 \sin 2 \delta_{23} \Rightarrow \sin 2 \delta_{23} = - \frac{q_1^2}{q_2^2} \sin 2\delta_{13}.
\end{equation} Taking \eqref{eq::diffd13}, expanding the sine of the difference and using \eqref{eq::relationsines} results in

\begin{equation}
\label{eq::auxcos}
    0 = \sin 2 \delta_{13} \left( 1 + q_1^2 \cos 2 \delta_{13} + q_2^2 \cos 2 \delta_{23} \right),
\end{equation} if one chooses to keep $\sin 2\delta_{13}$ by using \eqref{eq::relationsines}, and otherwise to

\begin{equation}
\label{eq::solvefor0}
    0 = \sin 2 \delta_{23} \left( 1 + q_2^2 \cos 2 \delta_{23} \pm \sqrt{q_1^4 - q_2^4 \sin^2 \delta_{23}}\right),
\end{equation} where we used $\cos 2\delta_{13} = \pm \sqrt{1 - \sin^2 \delta_{23}}$. Let us start by considering the second factor equal to zero. Some simple manipulations lead to the single solution

\begin{align}
\label{eq::x1lcos}
     \cos 2 \delta_{23} & = - \frac{1 - q_1^4 + q_2^4}{2 q_2^2}, & \cos 2 \delta_{13} & = - \frac{1 + q_1^4 - q_2^4}{2 q_1^2},
\end{align} where $\cos 2\delta_{13}$ was obtained by equating the second factor of \eqref{eq::auxcos} to zero. However, the existence of this solution is dependent on its boundedness. Demanding that both cosines be bounded leads to the simple constraint

\begin{equation}
    0 \leq \alpha_3^\prime \leq \frac{1}{4}.
\end{equation} We highlight here that the simplicity of this result reinforces the usefulness of the rotation introduced in \eqref{eq::rotationy3y8}.

Substituting this solution into \eqref{eq::x1leftapp} and making use of \eqref{eq::relationsines}, we obtain that

\begin{equation}
    x_1 = \frac{1}{4} - \left({\alpha_3^\prime}^ 2 + {\alpha_8^\prime}^2\right) = \frac{1}{4} - x_3.
\end{equation} This means that this solution must be the minimum of $x_1$, given the fact that within $\Gamma$ in this range of $\alpha_3^\prime$ we have obtained the smallest value of $x_1$ for a given $x_3$. If, however, $\alpha_3^\prime > \frac{1}{4}$ we turn to the first term in \eqref{eq::solvefor0}, which results in

\begin{equation}
    \delta_{23} = l \frac{\pi}{2},
\end{equation} forcing

\begin{equation}
    \delta_{13} = k \frac{\pi}{2},
\end{equation} with both $k$ and $l$ integers. Given the periodicity of $\pi$ of $x_1$ and our knowledge that $k = l = 0$ leads to the maximum of $x_1$, we are left with three possibilities,

\begin{align}
    (k, l) &= (1, 0): \quad x_1 = \frac{1}{4} - \left({\alpha_3^\prime}^ 2 + {\alpha_8^\prime}^2\right) + \frac{4}{3}\left({\alpha_3^\prime} - \frac{1}{4}\right)^2.\\
    (k, l) &= (0, 1): \quad x_1 = \frac{1}{3} \left(1 + 2 \alpha_3^\prime\right) \left(1 - \alpha_3^\prime - \sqrt{3} \alpha_8^\prime \right). \\
    (k, l) &= (1, 1): \quad x_1 = \frac{1}{3} \left(1 + 2 \alpha_3^\prime\right) \left(1 - \alpha_3^\prime + \sqrt{3} \alpha_8^\prime \right).
\end{align} The last possibility is immediately excluded due to $\alpha_8^\prime > 0$. Regarding the first two, some simple algebraic manipulations show that $(k,l) = (1, 0)$ is always smaller within the domain, for $\alpha_3^\prime > \frac{1}{4}$. Thus, the final result for the minimum of $x_1$ is the one shown in \eqref{eq::x1l}.


\begin{thebibliography}{9}

\bibitem{Lee:1973iz}
T.~D.~Lee,
Phys. Rev. D \textbf{8} (1973), 1226-1239
doi:10.1103/PhysRevD.8.1226

\bibitem{Branco:2011iw}
G.~C.~Branco, P.~M.~Ferreira, L.~Lavoura, M.~N.~Rebelo, M.~Sher and J.~P.~Silva,
Phys. Rept. \textbf{516} (2012), 1-102
doi:10.1016/j.physrep.2012.02.002
[arXiv:1106.0034 [hep-ph]].

\bibitem{Ivanov:2017dad}
I.~P.~Ivanov,
Prog. Part. Nucl. Phys. \textbf{95} (2017), 160-208
doi:10.1016/j.ppnp.2017.03.001
[arXiv:1702.03776 [hep-ph]].

\bibitem{Weinberg:1976hu}
S.~Weinberg,
Phys. Rev. Lett. \textbf{37} (1976), 657
doi:10.1103/PhysRevLett.37.657

\bibitem{Ivanov:2012ry}
I.~P.~Ivanov and E.~Vdovin,
Phys. Rev. D \textbf{86} (2012), 095030
doi:10.1103/PhysRevD.86.095030
[arXiv:1206.7108 [hep-ph]].

\bibitem{Darvishi:2021txa}
N.~Darvishi, M.~R.~Masouminia and A.~Pilaftsis,
Phys. Rev. D \textbf{104} (2021) no.11, 115017
doi:10.1103/PhysRevD.104.115017
[arXiv:2106.03159 [hep-ph]].

\bibitem{deMedeirosVarzielas:2021zqs}
I.~de Medeiros Varzielas, I.~P.~Ivanov and M.~Levy,
Eur. Phys. J. C \textbf{81} (2021) no.10, 918
doi:10.1140/epjc/s10052-021-09681-w
[arXiv:2107.08227 [hep-ph]].

\bibitem{deMedeirosVarzielas:2022kbj}
I.~de Medeiros Varzielas and D.~Ivo,
Eur. Phys. J. C \textbf{82} (2022) no.5, 415
doi:10.1140/epjc/s10052-022-10331-y
[arXiv:2202.00681 [hep-ph]].

\bibitem{Degee:2012sk}
A.~Degee, I.~P.~Ivanov and V.~Keus,
JHEP \textbf{02} (2013), 125
doi:10.1007/JHEP02(2013)125
[arXiv:1211.4989 [hep-ph]].

\bibitem{deMedeirosVarzielas:2017glw}
I.~de Medeiros Varzielas, S.~F.~King, C.~Luhn and T.~Neder,
Phys. Lett. B \textbf{775} (2017), 303-310
doi:10.1016/j.physletb.2017.11.005
[arXiv:1704.06322 [hep-ph]].

\bibitem{Ivanov:2014doa}
I.~P.~Ivanov and C.~C.~Nishi,
JHEP \textbf{01} (2015), 021
doi:10.1007/JHEP01(2015)021
[arXiv:1410.6139 [hep-ph]].

\bibitem{Ivanov:2020jra}
I.~P.~Ivanov and F.~Vaz\~ao,
JHEP \textbf{11} (2020), 104
doi:10.1007/JHEP11(2020)104
[arXiv:2006.00036 [hep-ph]].

\end{thebibliography}
\end{document}